\documentclass[10pt,twocolumn,twoside]{IEEEtran}

\usepackage{graphicx}
\usepackage{cite}
\usepackage[cmex10]{amsmath}
\usepackage{amssymb}
\usepackage{tikz}
\usepackage{pgfplots}
\pgfplotsset{compat=1.3}
\usetikzlibrary{arrows,shapes}
\DeclareMathAlphabet{\mathbit}{OML}{cmr}{bx}{it}
\DeclareMathOperator{\E}{E}

\DeclareMathOperator{\Variance}{var}
\DeclareMathOperator{\Mse}{mse}

\DeclareMathOperator{\Erf}{erf}
\DeclareMathOperator{\fieldR}{\mathbb{R}}

\newcommand{\ve}[1]{\boldsymbol{#1}}

\newcommand{\norm}[1]{\left|#1\right|}

\newcommand{\ex}[1]{\E \left[#1\right]}
\newcommand{\varout}[2]{\Variance_{#1}{\left(#2\right)}}
\newcommand{\mseout}[2]{\Mse_{#1}{\left(#2\right)}}
\renewcommand{\exp}[1]{{\rm e}^{#1}}

\newcommand{\erf}[1]{\Erf \left(#1\right)}

\title{A Pessimistic Approximation for\\the Fisher Information Measure}
\author{Manuel~S.~Stein and~Josef~A.~Nossek
\thanks{M. S. Stein is with the Digital Mathematics Group (DIMA), Mathematics Department (DWIS), Vrije Universiteit Brussel, Belgium (e-mail: manuel.stein@vub.ac.be). J. A. Nossek is with the Department of Teleinformatics Engineering, Universidade Federal do Cear\'{a}, Brasil, and with the Department of Electrical and Computer Engineering, Technische Universit\"at M\"unchen, Germany (e-mail: josef.a.nossek@tum.de).}
}
\begin{document}
\maketitle
\begin{abstract}
The problem of determining the intrinsic quality of a signal processing system with respect to the inference of an unknown deterministic parameter $\theta$ is considered. While the Fisher information measure $F(\theta)$ forms a classical tool for such a problem, direct computation of the information measure can become difficult in various situations. For the estimation theoretic performance analysis of nonlinear measurement systems, the form of the likelihood function can make the calculation of the information measure $F(\theta)$ challenging. In situations where no closed-form expression of the statistical system model is available, the analytical derivation of $F(\theta)$ is not possible at all. Based on the Cauchy-Schwarz inequality, we derive an alternative information measure $S(\theta)$. It provides a lower bound on the Fisher information $F(\theta)$ and has the property of being evaluated with the mean, the variance, the skewness and the kurtosis of the system model at hand. These entities usually exhibit good mathematical tractability or can be determined at low-complexity by real-world measurements in a calibrated setup. With various examples, we show that $S(\theta)$ provides a good conservative approximation for $F(\theta)$ and outline different estimation theoretic problems where the presented information bound turns out to be useful.
\end{abstract}
\begin{IEEEkeywords}
Cram\'er-Rao lower bound, estimation theory, Fisher information lower bound, smooth limiter, minimum Fisher information, nonlinear systems, squaring loss, worst-case noise.
\end{IEEEkeywords}
\IEEEpeerreviewmaketitle
\section{Introduction}
Suppose we are given a parametric system, characterized by a probability density or mass function $q(y;\theta)$, and face the problem of having to infer the deterministic but unknown system parameter $\theta\in\Theta$ from measurements at the system output $Y$. The output $Y$ takes random values $y\in\mathcal{Y}$, where $\mathcal{Y}$ denotes the support of the random variable $Y$. In such a situation, estimation theory \cite{Fisher22} \cite{Fisher25} provides a variety of useful tools. On the one hand, we have guidelines for the design of \emph{estimation algorithms} \cite{Kay93}, and on the other hand, corresponding \emph{performance bounds} \cite{Rao45} \cite{Cram46} \cite{Bhatt46} \cite{Bar49} \cite{Hamm50} \cite{Chap51}. While the latter have originally been derived to benchmark estimation algorithms, identify potential for further improvements, or to establish their efficiency, these error bounds have also become popular as a figure of merit for the design and optimization of the measurement system $q(y;\theta)$. Such a problem frequently arises in the field of signal processing, where not only the efficient extraction of information from noisy data is in the interest of engineers, but also the design of the physical measurement system $q(y;\theta)$ itself. Note that the layout of the measurement sensors can significantly influence technical properties such as computational complexity, power consumption, production cost, reliability, processing delay and system performance. Therefore, given the ability to modify the data gathering system $q(y;\theta)$ to an alternative design $p(z;\theta)$ with the altered output $Z$ exhibiting realizations $z\in\mathcal{Z}$, a rigorous method is required in order to draw a precise conclusion about the achievable performance of the two systems when operating with optimum estimation procedures $\hat{\theta}(\ve{y})$ or $\hat{\theta}(\ve{z})$. Note that here $\ve{y}\in\mathcal{Y}^N$ denotes a collection of $N$ independent realizations of the system outputs $Y$, such that
\begin{align}\label{definition:independence}
q(\ve{y};\theta)&=\prod_{n=1}^{N}q(y_n;\theta),\quad\quad\forall\ve{y}\in\mathcal{Y}^N.
\end{align}
\subsection{Estimation Theory and the Fisher Information Measure}
In order to motivate the use of an information measure, in the following, the performance of the estimator $\hat{\theta}(\ve{y})$ is analyzed. We restrict the discussion to the problem of performing unbiased estimation
\begin{align}
\int_{\mathcal{Y}^N} \hat{\theta}(\ve{y}) q(\ve{y};\theta) {\rm d} \ve{y} =\theta\label{def:unbiased}.
\end{align}
Further, we assume that the system $q(\ve{y};\theta)$ is differentiable in $\theta\in\Theta$ for every $\ve{y}\in\mathcal{Y}^N$, where the parameter set $\Theta$ is an open subset on the real line. All considered system models exhibit regularity, such that the statement
\begin{align}
\frac{\partial}{\partial \theta}\int_{\mathcal{Y}^N} f(\ve{y}) q(\ve{y};\theta) {\rm d} \ve{y}=\int_{\mathcal{Y}^N} f(\ve{y}) \frac{\partial q(\ve{y};\theta)}{\partial \theta} {\rm d} \ve{y}\label{def:regularity:n1}
\end{align}
holds for any function $f(\cdot)$ which does not present $\theta$ as an argument. Applying \eqref{def:regularity:n1} to \eqref{def:unbiased} we can set
\begin{align}
\int_{\mathcal{Y}^N} \hat{\theta}(\ve{y}) \frac{\partial q(\ve{y};\theta)}{\partial \theta}  {\rm d} \ve{y}&=1.\label{eq:diff:unbiased}
\end{align}
With the requirement
\begin{align}
\int_{\mathcal{Y}^N} q(\ve{y};\theta) {\rm d} \ve{y} =1,\quad\quad\forall\theta\in\Theta,
\end{align}
it follows that
\begin{align}\label{derivative:integral:pdf:zero}
\frac{\partial}{\partial \theta}\int_{\mathcal{Y}^N} q(\ve{y};\theta) {\rm d} \ve{y} =0,\quad\quad\forall\theta\in\Theta,
\end{align}
such that multiplying \eqref{derivative:integral:pdf:zero} by $\theta$ and expanding \eqref{eq:diff:unbiased}, we have
\begin{align}
\int_{\mathcal{Y}^N} (\hat{\theta}(\ve{y})-\theta) \frac{\partial q(\ve{y};\theta)}{\partial \theta}  {\rm d} \ve{y}&=1.\label{eq:diff:unbiased:expand}
\end{align}
Using the fact that
\begin{align}
\frac{\partial \ln q(\ve{y};\theta)}{\partial \theta}&=\frac{1 }{q(\ve{y};\theta)}\frac{\partial q(\ve{y};\theta)}{\partial \theta},
\end{align}
equation (\ref{eq:diff:unbiased:expand}) is manipulated, resulting in
\begin{align}
\int_{\mathcal{Y}^N} (\hat{\theta}(\ve{y})-\theta) \frac{\partial \ln q(\ve{y};\theta)}{\partial \theta} q(\ve{y};\theta) {\rm d} \ve{y}&=1.\label{eq:crlb:basis}
\end{align}
For two real-valued functions $f(\cdot)$, $g(\cdot)$ and a random variable $\ve{X}\in\fieldR^{N}$, the Cauchy-Schwarz inequality \cite{Koba11} states that
\begin{align}
&\int_{\mathcal{Y}^N} {f^2(\ve{x})} p(\ve{x}){\rm d}\ve{x} \int_{\mathcal{Y}^N} {g^2(\ve{x})} p(\ve{x}){\rm d}\ve{x}\notag\\
&\geq \bigg({\int_{\mathcal{Y}^N} f(\ve{x}) g(\ve{x}) p(\ve{x}){\rm d}\ve{x}}\bigg)^2,
\label{ineq:cauchy:schwarz}
\end{align}
where $p(\cdot)$ is a probability distribution function. By setting 
\begin{align}
f(\cdot)&=\hat{\theta}(\ve{y})-\theta,\\
 g(\cdot)&= \frac{\partial \ln q(\ve{y};\theta)}{\partial \theta}
\end{align}
and $p(\cdot)=q(\ve{y};\theta)$, this allows us to derive the inequality
\begin{align}
&\int_{\mathcal{Y}^N} (\hat{\theta}(\ve{y})-\theta)^2 q(\ve{y};\theta) {\rm d} \ve{y}\notag\\
&\geq\bigg(\int_{\mathcal{Y}^N} \Big( \frac{\partial \ln q(\ve{y};\theta)}{\partial \theta} \Big)^2 q(\ve{y};\theta) {\rm d} \ve{y}\bigg)^{-1}\label{eq:crlb:ineq}
\end{align}
from expression (\ref{eq:crlb:basis}). As long as the observations are independent \eqref{definition:independence} and each element $Y_n$ follows the identical statistical model
\begin{align}
q(y_n;\theta)=q(y;\theta),\quad\quad\forall n\in\{1,2,\ldots,N\},
\end{align}
the right hand side of (\ref{eq:crlb:ineq}) simplifies to
\begin{align}
&\int_{\mathcal{Y}^N} \bigg( \frac{\partial \ln q(\ve{y};\theta)}{\partial \theta} \bigg)^2 q(\ve{y};\theta) {\rm d} \ve{y} \notag\\
&=N \int_{\mathcal{Y}} \bigg( \frac{\partial \ln q({y};\theta)}{\partial \theta} \bigg)^2 q({y};\theta) {\rm d} {y}.
\end{align}
The left hand side of (\ref{eq:crlb:ineq}) is identified as the mean squared-error $\mseout{\ve{Y}}{\theta}$ of the estimator $\hat{\theta}(\ve{y})$, such that the Cram\'er-Rao inequality \cite{Rao45} \cite{Cram46}  for unbiased estimators 
\begin{align}
\mseout{\ve{Y}}{\theta}&=\varout{\ve{Y}}{\theta}\notag\\
&\geq\frac{1}{N F_{Y}(\theta)}\label{eq:crlb}
\end{align}
is obtained. Note that an estimator $\hat{\theta}(\ve{y})$, which asymptotically in $N$ attains equality with respect to \eqref{eq:crlb}, is called asymptotically efficient. Estimators designed along the principle of maximum-likelihood are known to exhibit efficiency in the asymptotic regime \cite[App. 7B]{Kay93} under mild conditions. Consequently, when $N$ is sufficiently large, the Fisher information
\begin{align}
F_Y(\theta)&=\int_{\mathcal{Y}} \bigg(\frac{\partial \ln{q(y;\theta)}}{\partial\theta}\bigg)^2 q(y;\theta)  {\rm d}y
\label{measure:fisher}
\end{align}
is a measure on the amount of intrinsic information about the unknown deterministic parameter $\theta$ contained in average within each observation of the random output $Y$. Note that the Fisher information measure also plays an important role for performance bounds in the Bayesian setting \cite{Trees68} \cite{Ziv69} \cite{Bobro75} \cite{Weiss85} \cite{Tichavsky98}, where the parameter $\theta$ is considered to be a random variable. A comprehensive overview on this topic, which is out of the scope of this article, can be found in \cite{Trees07}. 
\subsection{Relative Inference Capability}
As the inequality \eqref{eq:crlb} holds for all estimation procedures satisfying (\ref{def:unbiased}), the relative estimation theoretic quality of the modification $p(z;\theta)$ with respect to the reference system $q(y;\theta)$ can be assessed by the information ratio
\begin{align}\label{exact:loss:nonlinearity}
\chi(\theta)=\frac{F_Z(\theta)}{F_Y(\theta)}.
\end{align}
Note that $F_Z(\theta)$ is the Fisher information (\ref{measure:fisher}) evaluated on $\mathcal{Z}$ with respect to the parametric probability function $p(z;\theta)$.
\subsection{Fisher Information Lower Bound}
Using the information ratio \eqref{exact:loss:nonlinearity} for the design and the optimization of the measurement system $p(z;\theta)$ requires computing \eqref{measure:fisher} for the benchmark experiment $q(y;\theta)$ and all modifications $p(z;\theta)$ which are of interest. If due to the alteration the probability distribution $p(z;\theta)$ takes a complicated form, this can become difficult. In a situation where the parametric probabilistic model $p(z;\theta)$ which defines the statistical behavior of the random output $Z$ is unknown, a direct analytical formulation of the information measure (\ref{measure:fisher}) becomes impossible. However, if the mean
\begin{align}
\mu_1(\theta)&=\int_{\mathcal{Z}} z p_{z}(z;\theta) {\rm d}z
\end{align}
of the system output $Z$ and the variance
\begin{align}
\mu_2(\theta)&=\int_{\mathcal{Z}} \big( z-\mu_1(\theta) \big)^2 p_{z}(z;\theta) {\rm d}z
\end{align}
are known and are differentiable in $\theta$, it can be shown that the Fisher information $F(\theta)$ is in general bounded from below \cite{Sankaran64} \cite{Stein14}
\begin{align}
F(\theta)\geq \frac{1}{\mu_2(\theta)} \bigg( \frac{\partial \mu_1(\theta)}{\partial \theta} \bigg)^2.\label{eq:bound:unopt}
\end{align}
While in \cite{Stein14} the example of a hard-limited Gaussian model was given where the information bound \eqref{eq:bound:unopt} holds with equality, a simple counter example is immediately constructed. To this end, consider the system output to follow the generic parametric Gaussian distribution
\begin{align}
p(z;\theta)=\frac{1}{\sqrt{2\pi{\mu}_2(\theta)}} {\rm e}^{- \frac{(z-{\mu}_1(\theta))^2}{2{\mu}_2(\theta)} }.
\end{align}
The exact Fisher information is \cite[pp. 47]{Kay93}
\begin{align}
F(\theta)=\frac{1}{{\mu}_2(\theta)} \bigg( \frac{\partial {\mu}_1(\theta)}{\partial \theta} \bigg)^2 + \frac{1}{2{\mu}^2_2(\theta)} \bigg( \frac{\partial {\mu}_2(\theta)}{\partial \theta} \bigg)^2
\end{align}
and is equal to the right-hand side of \eqref{eq:bound:unopt} only for the special case where
\begin{align}
\frac{\partial {\mu}_2(\theta)}{\partial \theta}=0.\label{eq:second:const}
\end{align}
Obviously, the inequality (\ref{eq:bound:unopt}) does not take into account the contribution of the variance ${\mu}_2(\theta)$ to the Fisher information measure $F(\theta)$.
\subsection{Contribution and Outline}
Motivated by the insight obtained in the preceding section, we aim to improve the lower bound \eqref{eq:bound:unopt}. We achieve this by utilizing the Cauchy-Schwarz inequality (\ref{ineq:cauchy:schwarz}) with a more general approach than in \cite{Stein14} and subsequently maximizing the resulting expression. This leads to an alternative information measure $S(\theta)$, which forms a pessimistic approximation for $F(\theta)$ and exclusively contains the mean, the variance, the skewness, and the kurtosis of the system output model in parametric form. A discussion on situations where the derivative of the variance vanishes (\ref{eq:second:const}) shows that the inequality (\ref{eq:bound:unopt}) is contained in the presented result as a special case. Using various examples with continuous and discrete system outputs, we verify the quality of the alternative information measure $S(\theta)$. In order to demonstrate possible applications of the result and to provide further insights  through $S(\theta)$, we approximately determine the estimation theoretic information loss when squaring a standard Gaussian input distribution and advance the discussion concerning minimum Fisher information \cite{Stein14} \cite{Stam59} \cite{Boek77} \cite{UhrKli95} \cite{Zivo97} \cite{Berch09} \cite{Stoica11}. Finally, we mimic a situation of practical relevance. Measuring the output moments of a smooth limiting device with standard Gaussian input, we demonstrate how to conservatively establish the intrinsic inference capability $F(\theta)$ of a nonlinear signal processing system through the information measure $S(\theta)$, when the analytic form of the parametric output model $p(z;\theta)$ is not available.
\section{Improved Fisher Information Bound}
For the discussion, we additionally require the central output moments
\begin{align}
\mu_3(\theta)&=\int_{\mathcal{Z}} \big( z-\mu_1(\theta) \big)^3 p(z;\theta) {\rm d}z,\\
\mu_4(\theta)&=\int_{\mathcal{Z}} \big( z-\mu_1(\theta) \big)^4 p(z;\theta) {\rm d}z
\end{align}
and their normalized versions
\begin{align}
\bar{\mu}_3(\theta)&=\int_{\mathcal{Z}} \bigg( \frac{ z-\mu_1(\theta) } {\sqrt{\mu_2(\theta)}} \bigg)^3 p_{z}(z;\theta) {\rm d}z\notag\\
&={\mu_3(\theta)}{\mu_2^{-\frac{3}{2}}(\theta)},\\
\bar{\mu}_4(\theta)&=\int_{\mathcal{Z}} \bigg( \frac{ z-\mu_1(\theta) } {\sqrt{\mu_2(\theta)}} \bigg)^4 p_{z}(z;\theta) {\rm d}z\notag\\
&={\mu_4(\theta)}{\mu_2^{-2}(\theta)}.
\end{align}
Note that $\bar{\mu}_3(\theta)$ is refered to as the skewness, an indicator for the asymmetry of the output distribution $p(z;\theta)$, while $\bar{\mu}_4(\theta)$ is called the kurtosis, a characterization  for the shape of the output distribution $p(z;\theta)$. Both moments stand in relation via Pearson's inequality \cite{Pear16}
\begin{align}
\bar{\mu}_4(\theta)\geq\bar{\mu}_3^2(\theta)+1,\label{ineq:higher:moments}
\end{align}
for which a compact and elegant proof can be found in \cite{Wilk44}.
\subsection{Derivation of the Information Bound}
We apply the inequality \eqref{ineq:cauchy:schwarz} with
\begin{align}
f(z;\theta)&=\frac{\partial \ln{p(z;\theta)}}{\partial\theta}
\end{align} 
and
\begin{align}
g(z;\theta)= \bigg(\frac{z-\mu_1(\theta)}{\sqrt{\mu_2(\theta)}}\bigg) +\beta(\theta) \bigg(\frac{z-\mu_1(\theta)}{\sqrt{\mu_2(\theta)}}\bigg)^2 -\beta(\theta),
\end{align}
$\beta(\theta)\in\fieldR$, in order to derive a lower bound on the Fisher information
\begin{align}
F(\theta)=\int_{\mathcal{Z}} {f^2(z;\theta)} p(z;\theta){\rm d}z,\label{fisher:squared:form}
\end{align}
which takes into account the contribution of the variance to the Fisher information measure. With the manipulations
\begin{align}
&\int_{\mathcal{Z}}   \bigg(\frac{z-\mu_1(\theta)}{\sqrt{\mu_2(\theta)}}\bigg)\frac{\partial \ln{p_{z}(z;\theta)}}{\partial\theta} { p_{z}(z;\theta) } {\rm d}z=\notag\\
&=\frac{1}{\sqrt{\mu_2(\theta)}}  \Bigg(\int_{\mathcal{Z}}   z  \frac{\partial {p_{z}(z;\theta)}}{\partial\theta} {\rm d}z- \mu_1{(\theta)} \int_{\mathcal{Z}}   \frac{\partial {p_{z}(z;\theta)}}{\partial\theta} {\rm d}z \Bigg) \notag\\
&=\frac{1}{\sqrt{\mu_2(\theta)}}  \Bigg( \frac{\partial }{\partial\theta} \int_{\mathcal{Z}}   z  p_{z}(z;\theta) {\rm d}z- \mu_1{(\theta)}  \frac{\partial }{\partial\theta} \int_{\mathcal{Z}}  p_{z}(z;\theta) {\rm d}z \Bigg) \notag\\
&=\frac{1}{\sqrt{\mu_2(\theta)}} \frac{\partial \mu_1(\theta)}{\partial\theta} 
\end{align}
and
\begin{align}
&\int_{\mathcal{Z}}\bigg(\frac{z-\mu_1(\theta)}{\sqrt{\mu_2(\theta)}}\bigg)^2\frac{\partial \ln{p_{z}(z;\theta)}}{\partial\theta} { p_{z}(z;\theta) } {\rm d}z=\notag\\
&=\frac{1}{{\mu_2(\theta)}}  \Bigg(\int_{\mathcal{Z}}   z^2  \frac{\partial {p_{z}(z;\theta)}}{\partial\theta} {\rm d}z- 2 \mu_1{(\theta)} \int_{\mathcal{Z}}  z \frac{\partial {p_{z}(z;\theta)}}{\partial\theta} {\rm d}z \notag \\
&\quad\quad\quad\quad\quad\quad\quad\quad\quad\quad\quad\quad\quad\quad+  \mu_1^2{(\theta)} \int_{\mathcal{Z}}  \frac{\partial {p_{z}(z;\theta)}}{\partial\theta} {\rm d}z \Bigg) \notag\\
&=\frac{1}{{\mu_2(\theta)}}  \Bigg(  \frac{\partial }{\partial\theta} \int_{\mathcal{Z}}   z^2  {p_{z}(z;\theta)} {\rm d}z- 2 \mu_1{(\theta)}  \frac{\partial }{\partial\theta} \int_{\mathcal{Z}}  z {p_{z}(z;\theta)} {\rm d}z \Bigg) \notag\\
&=\frac{1}{{\mu_2(\theta)}}  \Bigg(  \frac{\partial }{\partial\theta} \big(\mu_2(\theta)+\mu_1^2(\theta)\big) - 2 \mu_1{(\theta)}  \frac{\partial \mu_1(\theta)}{\partial\theta}  \Bigg) \notag\\
&=\frac{1}{{\mu_2(\theta)}} \frac{\partial \mu_2(\theta)}{\partial\theta},
\end{align}
where we use the fact that
\begin{align}
\int_{\mathcal{Z}}  z^2  {p_{z}(z;\theta)} {\rm d}z = \mu_2(\theta)+\mu_1^2(\theta),
\end{align}
the identity
\begin{align}
&\int_{\mathcal{Z}} f(z;\theta) g(z;\theta) p(z;\theta){\rm d}z=\notag\\
&= \frac{1}{\sqrt{\mu_2(\theta)}} \frac{\partial \mu_1(\theta)}{\partial\theta} + \frac{\beta(\theta)}{{\mu_2(\theta)}} \frac{\partial \mu_2(\theta)}{\partial\theta},\label{approx:nominator}
\end{align}
is obtained. Note that
\begin{align}
&\int_{\mathcal{Z}} \beta(\theta)\frac{\partial \ln{p(z;\theta)}}{\partial\theta} { p(z;\theta) } {\rm d}z=\notag\\
&=\beta(\theta) \int_{\mathcal{Z}}\frac{\partial \ln{p(z;\theta)}}{\partial\theta} { p(z;\theta) } {\rm d}z\notag\\
&= \beta(\theta) \frac{\partial }{\partial\theta} \int_{\mathcal{Z}}  p(z;\theta)  {\rm d}z\notag\\
&=0.
\end{align}
Taking into account that
\begin{align}
\int_{\mathcal{Z}} \bigg(\frac{z-\mu_1(\theta)}{\sqrt{\mu_2(\theta)}}\bigg) { p(z;\theta) } {\rm d}z=0,\\
\int_{\mathcal{Z}} \bigg(\frac{z-\mu_1(\theta)}{\sqrt{\mu_2(\theta)}}\bigg)^2 { p(z;\theta) } {\rm d}z=1,
\end{align}
we get
\begin{align}
&\int_{\mathcal{Z}} g^2(z;\theta) p(z;\theta) {\rm d}z = \notag\\
&=1+ 2 \beta(\theta) \bar{\mu}_3(\theta) + \beta^2(\theta) \bar{\mu}_4(\theta)-\beta^2(\theta).\label{approx:denominator}
\end{align}
Therefore, from \eqref{ineq:cauchy:schwarz}, \eqref{fisher:squared:form}, \eqref{approx:nominator} and \eqref{approx:denominator} it can be shown that the Fisher information can in general not fall below
\begin{align}
F(\theta)&\geq \frac{\Big(\int_{\mathcal{Z}} f(z;\theta)g(z;\theta)p(z;\theta) {\rm d}z\Big)^2}{\int_{\mathcal{Z}}g^2(z;\theta)p(z;\theta) {\rm d}z}\notag\\
&=\frac{1}{{\mu_2(\theta)}} \frac{\Big(\frac{\partial \mu_1(\theta)}{\partial\theta} + \frac{\beta(\theta)}{\sqrt{\mu_2(\theta)}} \frac{\partial \mu_2(\theta)}{\partial\theta} \Big)^2}{1+ 2 \beta(\theta) \bar{\mu}_3(\theta) + \beta^2(\theta) (\bar{\mu}_4(\theta)-1)}\label{fisher:information:bound1}
\end{align}
for any $\beta(\theta)\in\fieldR$.
\subsection{Optimization of the Information Bound}
The factor $\beta(\theta)$ can be used to optimize the lower bound \eqref{fisher:information:bound1}. For the trivial choice $\beta(\theta)=0$, the expression  \eqref{fisher:information:bound1} degenerates to
\begin{align}
F(\theta)\geq \frac{1}{\mu_2(\theta)} \bigg(\frac{\partial \mu_1(\theta)}{\partial\theta} \bigg)^2,
\end{align}
which turns out to be the bound in \eqref{eq:bound:unopt}. In order to improve this result, note that the problem
\begin{align}
x^\star=\arg \max_{x\in\fieldR} h(x)
\end{align}
with
\begin{align}
h(x)=\frac{(a+x b)^2}{1 + 2x c + x^2 d},
\end{align}
and $bc-ad\neq0$ has a unique maximizing solution
\begin{align}
x^\star=\frac{ac-b}{bc-ad}.
\end{align}
Consequently, the tightest form of (\ref{fisher:information:bound1}) is given by 
\begin{align}\label{fisher:information:bound:definition}
F(\theta)&\geq S(\theta),
\end{align}
where
\begin{align}\label{fisher:information:bound:opt}
S(\theta)=\frac{1}{{\mu_2(\theta)}} \frac{\Big(\frac{\partial \mu_1(\theta)}{\partial\theta} + \frac{\beta^\star(\theta)}{\sqrt{\mu_2(\theta)}} \frac{\partial \mu_2(\theta)}{\partial\theta} \Big)^2}{1+ 2 \beta^\star(\theta) \bar{\mu}_3(\theta) + {\beta^\star}^2(\theta) (\bar{\mu}_4(\theta)-1)}
\end{align}
with the optimization result
\begin{align}
\beta^\star(\theta)&=\frac{\frac{\partial \mu_1(\theta)}{\partial\theta} \bar{\mu}_3 (\theta) - 
\frac{1}{\sqrt{\mu_2(\theta)}} \frac{\partial \mu_2(\theta)}{\partial\theta}}
{\frac{1}{\sqrt{\mu_2(\theta)}} \frac{\partial \mu_2(\theta)}{\partial\theta} \bar{\mu}_3 (\theta) -\frac{\partial \mu_1(\theta)}{\partial\theta} (\bar{\mu}_4 (\theta)-1)}\notag\\
&=\frac{\frac{\partial \mu_1(\theta)}{\partial\theta} \sqrt{\mu_2(\theta)}\bar{\mu}_3 (\theta) - \frac{\partial \mu_2(\theta)}{\partial\theta}}
{ \frac{\partial \mu_2(\theta)}{\partial\theta} \bar{\mu}_3 (\theta) -\frac{\partial \mu_1(\theta)}{\partial\theta} \sqrt{\mu_2(\theta)} (\bar{\mu}_4 (\theta)-1)}.
\end{align}
Note that for the case where it holds that
\begin{align}\label{simplifying:charactristic}
\frac{\partial \mu_1(\theta)}{\partial\theta} \sqrt{\mu_2(\theta)} \bar{\mu}_3 (\theta)=\frac{\partial \mu_2(\theta)}{\partial\theta},
\end{align}
the optimization of \eqref{fisher:information:bound:opt} results in 
\begin{align}
\beta^\star(\theta)=0
\end{align}
and the approximation obtains the compact form \eqref{eq:bound:unopt}.
The inequality (\ref{fisher:information:bound:definition}) states that the derived information measure $S(\theta)$ is always dominated by the Fisher information measure $F(\theta)$. Therefore, $S(\theta)$ gives a cautious approximation for $F(\theta)$. The Fisher information $F(\theta)$ requires integrating the squared score function
\begin{align}
f^2(z;\theta)&=\bigg(\frac{\partial \ln{p(z;\theta)}}{\partial\theta}\bigg)^2,
\end{align}
while in contrast, the alternative information measure $S(\theta)$ exclusively needs the mean $\mu_1(\theta)$, the variance $\mu_2(\theta)$, the skewness $\bar{\mu}_3(\theta)$, and the kurtosis $\bar{\mu}_4(\theta)$ in parametric form. These entities are usually analytically tractable, can be determined by simple measurements in a calibrated setup and are well studied for various probability laws. Note that based on raw moments and cumulants, an alternative to the bound \eqref{fisher:information:bound:definition} is found in \cite{Jarrett84}.
\subsection{Positiveness of the Information Bound}
To ensure that the approximation \eqref{fisher:information:bound:opt} is always positive, it has to hold that
\begin{align}
1+ 2 \beta(\theta) \bar{\mu}_3(\theta) + \beta^2(\theta) (\bar{\mu}_4(\theta)-1)\geq 0, \quad\quad \forall \beta(\theta).
\end{align}
In order to demonstrate that this is the case, consider the fact that by construction
\begin{align}\label{inequality:denominator:bound:initial}
&\big(1+\beta(\theta)\sqrt{\bar{\mu}_4(\theta)-1}\big)^2\notag\\
&=1+ 2 \beta(\theta) \sqrt{\bar{\mu}_4(\theta)-1} + \beta^2(\theta) (\bar{\mu}_4(\theta)-1) \notag\\
&\geq 0, \quad\quad \forall \beta(\theta).
\end{align}
With Pearson's inequality \eqref{ineq:higher:moments}, we have
\begin{align}
\sqrt{\bar{\mu}_4(\theta)-1}\geq \left| \bar{\mu}_3(\theta) \right|, \quad\quad \forall \theta,
\end{align}
such that with \eqref{inequality:denominator:bound:initial} the inequality
\begin{align}\label{inequality:denominator:bound}
1+ 2 \beta(\theta) \left| \bar{\mu}_3(\theta) \right| + \beta^2(\theta) (\bar{\mu}_4(\theta)-1) &\geq 0, \quad \forall \beta(\theta)
\end{align}
is obtained. As \eqref{inequality:denominator:bound} holds irrespectively if $\beta(\theta)$ is positive or negative, we equivalently have
\begin{align}
1+ 2 \beta(\theta) \bar{\mu}_3(\theta) + \beta^2(\theta) (\bar{\mu}_4(\theta)-1) &\geq 0, \quad \forall \beta(\theta)
\end{align}
and the information bound $S(\theta)$ is always positive.
\section{Special Cases of the Information Bound}
In order to derive simplified forms of the derived information measure \eqref{fisher:information:bound:opt}, let us consider some special cases in the following.
\subsection{Constant Mean}
For the situation where the mean $\mu_1(\theta)$ does not vary with the system parameter $\theta$, i.e.,
\begin{align}
\frac{\partial \mu_1(\theta)}{\partial\theta}&=0,\quad\quad\forall\theta\in\Theta,
\end{align}
we obtain
\begin{align}
\beta^\star(\theta)=-\frac{1}{\bar{\mu}_3 (\theta)},
\end{align}
such that a pessimistic approximation for $F(\theta)$ is
\begin{align}\label{bound:const:first}
S(\theta) &= \frac{1}{{\mu_2(\theta)}} \frac{\Big(- \frac{1}{\bar{\mu}_3 (\theta) \sqrt{\mu_2(\theta)}} \frac{\partial \mu_2(\theta)}{\partial\theta} \Big)^2}{1 - 2  + \frac{ (\bar{\mu}_4(\theta)-1)}{\bar{\mu}^2_3 (\theta)}}\notag\\
&=\frac{1}{  {\mu^2_2(\theta)}} \frac{\Big(\frac{\partial \mu_2(\theta)}{\partial\theta} \Big)^2}{\bar{\mu}_4(\theta)  - \bar{\mu}^2_3 (\theta)-1}.
\end{align}
\subsection{Constant Variance}
When the variance $\mu_2(\theta)$ is constant with respect to $\theta$, i.e.,
\begin{align}
\frac{\partial \mu_2(\theta)}{\partial\theta}&=0,\quad\quad\forall\theta\in\Theta,
\end{align}
it holds that
\begin{align}
\beta^\star(\theta)&=- \frac{\bar{\mu}_3 (\theta)}{(\bar{\mu}_4 (\theta)-1)}.
\end{align}
In this situation
\begin{align}
S(\theta) &= \frac{1}{\mu_2(\theta)} \frac{\Big( \frac{\partial \mu_1(\theta)}{\partial\theta} \Big)^2}
{1 - 2 \frac{  \bar{\mu}_3^2(\theta) }{ (\bar{\mu}_4(\theta)-1)} + \frac{  \bar{\mu}_3^2(\theta) }{ (\bar{\mu}_4(\theta)-1)  } } \notag\\
&= \frac{1}{\mu_2(\theta)} \frac{\Big( \frac{\partial \mu_1(\theta)}{\partial\theta} \Big)^2}
{1 -  \frac{  \bar{\mu}_3^2(\theta) }{ (\bar{\mu}_4(\theta)-1)  } }.\label{bound:const:second}
\end{align}
Note that (\ref{bound:const:second}) equals the expression in \eqref{eq:bound:unopt} whenever the skewness $\bar{\mu}_3(\theta)$ vanishes. In general, the relation \eqref{ineq:higher:moments} between skewness and kurtosis makes \eqref{bound:const:second} larger than the unoptimized bound \eqref{eq:bound:unopt}.
\subsection{Symmetric Probability Distributions}
For symmetric output distributions with zero skewness, i.e.,
\begin{align}
\bar{\mu}_3(\theta)=0, 
\end{align}
we verify that the optimization of the information bound derived in (\ref{fisher:information:bound:opt}) results in
\begin{align}
\beta^\star(\theta)=\frac{ \frac{\partial \mu_2(\theta)}{\partial\theta}}
{  \frac{\partial \mu_1(\theta)}{\partial\theta} \sqrt{\mu_2(\theta)} (\bar{\mu}_4 (\theta)-1)},
\end{align}
such that
\begin{align}
S(\theta) &= \frac{1}{{\mu_2(\theta)}} \frac{\Big(\frac{\partial \mu_1(\theta)}{\partial\theta} + \frac{ \big( \frac{\partial \mu_2(\theta)}{\partial\theta} \big)^2 }{\frac{\partial \mu_1(\theta)}{\partial\theta} {\mu_2(\theta)} (\bar{\mu}_4 (\theta)-1)} \Big)^2}{1 + \Big(\frac{ \frac{\partial \mu_2(\theta)}{\partial\theta}}
{  \frac{\partial \mu_1(\theta)}{\partial\theta} \sqrt{\mu_2(\theta)} (\bar{\mu}_4 (\theta)-1)}\Big)^2 (\bar{\mu}_4(\theta)-1)}\notag\\
 &= \frac{\Big(\frac{\partial \mu_1(\theta)}{\partial\theta}\Big)^2 {\mu_2(\theta)} (\bar{\mu}_4 (\theta)-1) + { \big( \frac{\partial \mu_2(\theta)}{\partial\theta} \big)^2 } }{{\mu^2_2(\theta)} (\bar{\mu}_4(\theta)-1)} 
\notag\\
 &= \frac{1}{\mu_2(\theta)} \bigg(\frac{\partial \mu_1(\theta)}{\partial\theta}\bigg)^2 + \frac{1}{\mu^2_2(\theta) (\bar{\mu}_4 (\theta)-1) } \bigg( \frac{\partial \mu_2(\theta)}{\partial\theta} \bigg)^2.\label{bound:simple:skewness}
\end{align}
\section{Approximation Quality - Continuous Outputs}
In order to demonstrate the tightness of the derived information bound $S(\theta)$, we use different examples where $F(\theta)$ can be derived in a compact form. First, we discuss several well-studied distributions with continuous support $\mathcal{Z}$. 
\subsection{Gaussian System Model}
Consider the system output $Z$ which follows a generic Gaussian distribution of parametric form
\begin{align}
p(z;\theta)=\frac{1}{\sqrt{2\pi{\nu}_2(\theta)}} {\rm e}^{- \frac{(z-{\nu}_1(\theta))^2}{2{\nu}_2(\theta)} }.
\end{align}
The exact Fisher information measure is given by \cite[p. 47]{Kay93}
\begin{align}\label{exact:fisher:gaussian}
F(\theta)=\frac{1}{{\nu}_2(\theta)} \bigg( \frac{\partial {\nu}_1(\theta)}{\partial \theta} \bigg)^2 + \frac{1}{2{\nu}^2_2(\theta)} \bigg( \frac{\partial {\nu}_2(\theta)}{\partial \theta} \bigg)^2.
\end{align}
As the mean, variance, skewness, and kurtosis of $p(z;\theta)$ are
\begin{align}
{\mu}_1 (\theta)&={\nu}_1(\theta),\\
{\mu}_2 (\theta)&={\nu}_2(\theta),\\
\bar{\mu}_3 (\theta)&=0,\\
\bar{\mu}_4 (\theta)&=3,
\end{align}
with \eqref{bound:simple:skewness} we get the approximation
\begin{align}\label{approximated:fisher:gaussian}
S(\theta) &= \frac{1}{\mu_2(\theta)} \bigg(\frac{\partial \mu_1(\theta)}{\partial\theta}\bigg)^2 + \frac{1}{\mu^2_2(\theta) (\bar{\mu}_4 (\theta)-1) } \bigg( \frac{\partial \mu_2(\theta)}{\partial\theta} \bigg)^2\notag\\
&=  \frac{1}{\nu_2(\theta)} \bigg(\frac{\partial \nu_1(\theta)}{\partial\theta}\bigg)^2 + \frac{1}{2 \nu^2_2(\theta) } \bigg( \frac{\partial \nu_2(\theta)}{\partial\theta} \bigg)^2.
\end{align}
Comparing \eqref{approximated:fisher:gaussian} to the original information measure in \eqref{exact:fisher:gaussian} it is obvious that here $S(\theta)$ forms a tight lower bound for $F(\theta)$.
\subsection{Exponential System Model}
As another example, we analyze the situation where samples from a parametric exponential distribution 
\begin{align}
p(z;\theta)=\nu(\theta) {\rm e}^{-\nu(\theta) z },
\end{align}
with $\nu(\theta)>0$ and $z\geq0$ can be collected at the random system output $Z$. The score function under this model is
\begin{align}
\frac{\partial \ln p(z;\theta)}{\partial \theta}=\frac{1}{\nu(\theta)} \frac{\partial \nu(\theta)}{\partial \theta} - z \frac{\partial \nu(\theta)}{\partial \theta},
\end{align}
such that the Fisher information is evaluated to be
\begin{align}
F(\theta)&=
\int_{\mathcal{Z}} \bigg(\frac{\partial \ln{p(z;\theta)}}{\partial\theta}\bigg)^2 p(z;\theta)  {\rm{d}}z\notag\\
&=\frac{1}{\nu^2(\theta)} \bigg(\frac{\partial \nu(\theta)}{\partial \theta}\bigg)^2 + \bigg(\frac{\partial \nu(\theta)}{\partial \theta}\bigg)^2  \int_{\mathcal{Z}} z^2p(z;\theta)  {\rm{d}}z\notag\\
&- \frac{2}{\nu(\theta)} \bigg(\frac{\partial \nu(\theta)}{\partial \theta}\bigg)^2 \int_{\mathcal{Z}} z p(z;\theta)  {\rm{d}}z\notag\\
&=\frac{1}{\nu^2(\theta)} \bigg(\frac{\partial \nu(\theta)}{\partial \theta}\bigg)^2\label{fisher:exponential}
\end{align}
by using
\begin{align}
\int_{\mathcal{Z}} z p(z;\theta)  {\rm{d}}z&=\frac{1}{\nu(\theta)},\\
\int_{\mathcal{Z}} z^2 p(z;\theta)  {\rm{d}}z&=\frac{2}{\nu^2(\theta)}.
\end{align}
For the approximation $S(\theta)$ the required mean, the variance, the skewness, and the kurtosis are
\begin{align}
{\mu}_1 (\theta)&=\frac{1}{\nu (\theta)},\\
{\mu}_2 (\theta)&=\frac{1}{\nu^2 (\theta)},\\
\bar{\mu}_3 (\theta)&=2,\\
\bar{\mu}_4 (\theta)&=3,
\end{align}
such that
\begin{align}
\frac{\partial \mu_1(\theta)}{\partial\theta} \sqrt{\mu_2(\theta)} \bar{\mu}_3 (\theta)&=-\frac{2}{\nu^3(\theta)}\frac{\partial \nu(\theta)}{\partial \theta}\notag\\
&=\frac{\partial \mu_2(\theta)}{\partial\theta},
\end{align}
producing the optimization result $\beta^\star(\theta)=0$ as noted in \eqref{simplifying:charactristic}. The approximation is therefore given by the simplified form
\begin{align}
S(\theta) &= \frac{1}{\mu_2(\theta)} \bigg(\frac{\partial \mu_1(\theta)}{\partial\theta}\bigg)^2\notag\\
&=\nu^2 (\theta) \bigg(- \frac{1}{\nu^2(\theta)}\frac{\partial \nu(\theta)}{\partial\theta}\bigg)^2\notag\\
&= \frac{1}{\nu^2(\theta)} \bigg(\frac{\partial \nu(\theta)}{\partial\theta}\bigg)^2,
\end{align}
which matches the Fisher information $F(\theta)$ in (\ref{fisher:exponential}) exactly.
\subsection{Laplacian System Model}
For a third example, we assume that the output $Z$ follows a parametric Laplace distribution with zero mean, i.e.,
\begin{align}
p(z;\theta)=\frac{1}{2\nu(\theta)} {\rm e}^{-\frac{\norm{z}}{\nu(\theta)} },
\end{align}
with $\nu(\theta)>0$. The score function is given by
\begin{align}
\frac{\partial \ln p(z;\theta)}{\partial \theta}=-\frac{1}{\nu(\theta)} \frac{\partial \nu(\theta)}{\partial \theta} + \frac{\norm{z}}{\nu^2(\theta)} \frac{\partial \nu(\theta)}{\partial \theta}
\end{align}
and the exact Fisher information is found to be
\begin{align}
F(\theta)&=
\int_{\mathcal{Z}} \bigg(\frac{\partial \ln{p(z;\theta)}}{\partial\theta}\bigg)^2 p_{z}(z;\theta)  {\rm{d}}z\notag\\
&=\frac{1}{\nu^2(\theta)} \bigg(\frac{\partial \nu(\theta)}{\partial \theta}\bigg)^2.
\end{align}
The first four moments of the output $Z$ are
\begin{align}
{\mu}_1 (\theta)&=0,\\
{\mu}_2 (\theta)&=2\nu^2 (\theta),\\
\bar{\mu}_3 (\theta)&=0,\\
\bar{\mu}_4 (\theta)&=6.
\end{align}
As the first moment is constant with respect to the system parameter $\theta$, the approximation takes the form \eqref{bound:const:first}
\begin{align}
S(\theta) &=\frac{1}{  {\mu^2_2(\theta)}} \frac{\Big(\frac{\partial \mu_2(\theta)}{\partial\theta} \Big)^2}{(\bar{\mu}_4(\theta)-1)}\notag\\
&=\frac{1}{  4\nu^4 (\theta)} \frac{\Big(4\nu(\theta)\frac{\partial \nu(\theta)}{\partial\theta} \Big)^2}{5}\notag\\
&=\frac{4}{  5} \frac{1}{\nu^2 (\theta)} \bigg(\frac{\partial \nu(\theta)}{\partial\theta} \bigg)^2.
\end{align}
In contrast to the other examples, the information bound $S(\theta)$ is loose under the Laplacian system model.
\section{Approximation Quality - Discrete Outputs}
In the following, we extend the discussion on the tightness of $S(\theta)$ to the case where the system output $Z$ takes values from a discrete alphabet $\mathcal{Z}$.
\subsection{Bernoulli System Model}
As a first example for such kind of system outputs, observations from a parametric Bernoulli distribution with
\begin{align}
p(z=1;\theta)&=1-p(z=0;\theta)\notag\\
&=\nu(\theta)
\end{align}
are considered, where $0<\nu(\theta)<1, \forall\theta\in\Theta$. The Fisher information measure under this model is
\begin{align}
F(\theta)&=\int_{\mathcal{Z}} \bigg(\frac{\partial \ln{p(z;\theta)}}{\partial\theta}\bigg)^2 p(z;\theta)  {\rm{d}}z\notag\\
&=\sum_{\mathcal{Z}} \bigg(\frac{\partial {p(z;\theta)}}{\partial\theta}\bigg)^2 \frac{1}{p(z;\theta)}\notag\\
&=\frac{\big(\frac{\partial {p(z=1;\theta)}}{\partial\theta}\big)^2}{p(z=1;\theta)}+\frac{\big(\frac{\partial {p(z=0;\theta)}}{\partial\theta}\big)^2}{p(z=0;\theta)}\notag\\
&=\frac{1}{\nu(\theta)(1-\nu(\theta))} \bigg(\frac{\partial \nu(\theta)}{\partial \theta}\bigg)^2.\label{exact:fisher:bernoulli}
\end{align}
The mean and the variance are
\begin{align}
\mu_1(\theta)&=\nu(\theta),\\
\mu_2(\theta)&=\nu(\theta)(1-\nu(\theta)),
\end{align}
with their derivatives
\begin{align}
\frac{\partial\mu_1(\theta)}{\partial\theta}&=\frac{\partial \nu(\theta)}{\partial \theta},\\
\frac{\partial\mu_2(\theta)}{\partial\theta}&=\big(1-2\nu(\theta)\big)\frac{\partial \nu(\theta)}{\partial \theta}.
\end{align}
The skewness is
\begin{align}
\bar{\mu}_3(\theta)&=\sum_{\mathcal{Z}}  \bigg(\frac{z-\mu_1(\theta)}{\sqrt{\mu_2(\theta)}} \bigg)^3 p(z;\theta) \notag\\
&= \frac{1-2\nu(\theta)}{\sqrt{\nu(\theta)(1-\nu(\theta))}}
\end{align}
and the kurtosis
\begin{align}
\bar{\mu}_4(\theta)&=\sum_{\mathcal{Z}}  \bigg(\frac{z-\mu_1(\theta)}{\sqrt{\mu_2(\theta)}} \bigg)^4 p(z;\theta) \notag\\
&=\frac{1}{\nu(\theta)(1-\nu(\theta))} - 3.
\end{align}
As
\begin{align}
\frac{\partial \mu_1(\theta)}{\partial\theta} \sqrt{\mu_2(\theta)} \bar{\mu}_3 (\theta)&={\big(1-2\nu(\theta)\big)} \frac{\partial \nu(\theta)}{\partial \theta} \notag\\
&=\frac{\partial \mu_2(\theta)}{\partial\theta}
\end{align}
and consequently the optimization of the information bound results in $\beta^\star(\theta)=0$, the approximation takes its simplified form \eqref{eq:bound:unopt}
\begin{align}
S(\theta)&= \frac{1}{\mu_2(\theta)} \bigg(\frac{\partial \mu_1(\theta)}{\partial\theta}\bigg)^2\notag\\
&=\frac{1}{\nu(\theta)(1-\nu(\theta))}\bigg(\frac{\partial \nu(\theta)}{\partial \theta}\bigg)^2.
\end{align}
It becomes clear that for a binary system output $Z$ following a parametric Bernoulli distribution, the derived expression $S(\theta)$ is a tight approximation for the original inference capability $F(\theta)$ given in \eqref{exact:fisher:bernoulli}.
\subsection{Poisson System Model}
As a second example with discrete output, we consider the Poisson distribution. The samples $z$ at the output $Z$ are distributed according to the model
\begin{align}
p(z;\theta)=\frac{\nu^z(\theta)}{z!} {\rm e}^{-\nu(\theta)},
\end{align}
with $\mathcal{Z}=\{0,1,2,\ldots\}$ and $\nu(\theta)>0, \forall\theta\in\Theta$.
The derivative of the log-likelihood is given by
\begin{align}
\frac{\partial \ln p(z;\theta)}{\partial \theta}=\frac{z}{\nu(\theta)} \frac{\partial{\nu(\theta)}}{\partial\theta}-\frac{\partial{\nu(\theta)}}{\partial\theta},
\end{align}
such that we calculate
\begin{align}\label{fisher:information:poisson:exact}
F(\theta)&=\int_{\mathcal{Z}} \bigg(\frac{\partial \ln{p(z;\theta)}}{\partial\theta}\bigg)^2 p(z;\theta)  {\rm{d}}z\notag\\
&=\int_{\mathcal{Z}} \bigg(\frac{z}{\nu(\theta)} \frac{\partial{\nu(\theta)}}{\partial\theta}-\frac{\partial{\nu(\theta)}}{\partial\theta}\bigg)^2 p(z;\theta)  {\rm{d}}z\notag\\
&=\frac{1}{\nu^2(\theta)}\bigg( \frac{\partial{\nu(\theta)}}{\partial\theta}\bigg)^2 \int_{\mathcal{Z}} z^2 p(z;\theta) {\rm{d}}z\notag\\
&-\frac{2}{\nu(\theta)} \bigg( \frac{\partial{\nu(\theta)}}{\partial\theta}\bigg)^2 \int_{\mathcal{Z}} z p(z;\theta) {\rm{d}}z
+\bigg( \frac{\partial{\nu(\theta)}}{\partial\theta}\bigg)^2\notag\\
&=\frac{1}{\nu(\theta)}\bigg( \frac{\partial{\nu(\theta)}}{\partial\theta}\bigg)^2,
\end{align}
where we have used
\begin{align}
\int_{\mathcal{Z}} z p(z;\theta) {\rm{d}}z &= \sum_{z=0}^{\infty} z \frac{\nu^z(\theta)}{z!} {\rm e}^{-\nu(\theta)}\notag\\
&=\nu(\theta),\\
\int_{\mathcal{Z}} z^2 p(z;\theta) {\rm{d}}z &=\sum_{z=0}^{\infty} z^2 \frac{\nu^z(\theta)}{z!} {\rm e}^{-\nu(\theta)}\notag\\
&=\nu(\theta)+\nu^2(\theta).
\end{align}
In order to apply the approximation $S(\theta)$, we require the mean and the variance
\begin{align}
{\mu}_1 (\theta)={\mu}_2 (\theta)=\nu(\theta),
\end{align}
the skewness and the kurtosis
\begin{align}
\bar{\mu}_3 (\theta)&=\frac{1}{\sqrt{\nu(\theta)}},\\
\bar{\mu}_4 (\theta)&=\frac{1}{\nu(\theta)}+3.
\end{align}
As these quantities exhibit the property
\begin{align}
\frac{\partial \mu_1(\theta)}{\partial\theta} \sqrt{\mu_2(\theta)} \bar{\mu}_3 (\theta)&= \frac{\partial \nu(\theta)}{\partial \theta} \notag\\
&=\frac{\partial \mu_2(\theta)}{\partial\theta},
\end{align}
we obtain $\beta^\star(\theta)=0$ and the approximation for this example 
\begin{align}
S(\theta)&= \frac{1}{\mu_2(\theta)} \Big(\frac{\partial \mu_1(\theta)}{\partial\theta}\Big)^2\notag\\
&=\frac{1}{\nu(\theta)}\Big(\frac{\partial \nu(\theta)}{\partial \theta}\Big)^2
\end{align}
is tight when comparing it to \eqref{fisher:information:poisson:exact}.
\section{Applications of the Information Bound}
Finally, we want to outline possible applications of the presented result and the opportunities provided by an information bound like \eqref{fisher:information:bound:definition}. To this end, we present three problems for which $S(\theta)$ provides interesting and useful insights. The problems discussed cover theoretic as well as practical aspects in statistical signal processing.
\subsection{Worst-Case Noise and Minimum Fisher Information}
An important question in signal processing is to specify the worst-case noise distribution under the considered system model \cite{Huber72}. A common assumption in the field is that noise affects technical receive systems in an additive way. Therefore a model of high practical relevance is 
\begin{align}
Z=x(\theta)+W,\label{def:system:additive}
\end{align}
where $x(\theta)$ is a deterministic pilot signal modulated by the unknown parameter $\theta$ (for example attenuation, time-delay, frequency-offset, etc.) and $W$ is additive independent random noise with zero mean
\begin{align}
\ex{W}=0.
\end{align}
If in addition the noise has the property
\begin{align}
\ex{W^2}=\nu,
\end{align}
i.e., the second central moment of $Z$ is constant, it is well-understood that assuming the noise component $W$ to follow the Gaussian probability density function
\begin{align}
p(w)=\frac{1}{\sqrt{2\pi{\nu}}} {\rm e}^{- \frac{w^2}{2{\nu}} },
\end{align}
leads to minimum Fisher information $F(\theta)$ \cite{Stam59} \cite{Boek77}. Therefore, under an estimation theoretic perspective, Gaussian noise is the worst-case assumption for an additive system like (\ref{def:system:additive}) with constant second output moment \cite{Stoica11}. The presented bounding approach \eqref{fisher:information:bound:definition} allows us to generalize these statements. If for any system $p(z;\theta)$ (including non-additive systems) the output $Z$ exhibits the characteristic 
\begin{align}
\mu_1(\theta)&=\ex{Z}\notag\\
&=x(\theta),\label{rest:gauss:first:para}\\
\mu_2(\theta)&=\ex{\big(Z-\mu_1(\theta)\big)^2}\notag\\
&=\nu,\label{rest:gauss:second:const}
\end{align}
the presented result \eqref{bound:const:second} shows that $F(\theta)$ cannot violate
\begin{align}
F(\theta) &\geq \frac{1}{\mu_2(\theta)} \frac{\Big( \frac{\partial \mu_1(\theta)}{\partial\theta} \Big)^2}
{1 -  \frac{  \bar{\mu}_3^2(\theta) }{ (\bar{\mu}_4(\theta)-1)  } }.
\end{align}
This lower bound is minimized by a symmetric distribution, i.e., $\bar{\mu}_3(\theta)=0$. The resulting expression
\begin{align}
F(\theta) &\geq \frac{1}{\mu_2(\theta)} \bigg( \frac{\partial \mu_1(\theta)}{\partial\theta} \bigg)^2,
\end{align}
reaches equality under an additive Gaussian system model
\begin{align}
p(z;\theta)=\frac{1}{\sqrt{2\pi{\nu}}} {\rm e}^{- \frac{(z-x(\theta))^2}{2{\nu}} },
\end{align}
such that the worst-case model assumption with respect to Fisher information under the considered restrictions (\ref{rest:gauss:first:para}) and (\ref{rest:gauss:second:const}) is, in general, additive and Gaussian.
In the more general setting, where also the output variance exhibits a dependency on the system parameter $\theta$,
\begin{align}
\mu_1(\theta)&=\ex{Z}\notag\\
&=x(\theta),\\
\mu_2(\theta)&=\ex{(Z-\mu_1(\theta))^2}\notag\\
&=\nu(\theta)
\end{align}
and additionally the output distribution is symmetric, i.e.,
\begin{align}
\bar{\mu}_3(\theta)&=0,
\end{align}
the presented result allows us to conclude that the Fisher information is in general bounded from below by 
\begin{align}\label{lower:bound:nonadditive:gauss}
F(\theta)&\geq \frac{1}{\nu(\theta)} \bigg(\frac{\partial x(\theta)}{\partial\theta}\bigg)^2 + \frac{1}{\nu^2(\theta) (\bar{\mu}_4 (\theta)-1) } \bigg( \frac{\partial \nu(\theta)}{\partial\theta} \bigg)^2.
\end{align}
As the system model
\begin{align}
p(z;\theta)=\frac{1}{\sqrt{2\pi{\nu}(\theta)}} {\rm e}^{- \frac{(z-x(\theta))^2}{2{\nu}(\theta)} }\label{model:Gaussian:generic:signal}
\end{align}
exhibits the inference capability
\begin{align}\label{fisher:information:additive:gauss}
F(\theta)&=\frac{1}{\nu(\theta)} \bigg(\frac{\partial x(\theta)}{\partial\theta}\bigg)^2 + \frac{1}{2 \nu^2(\theta)} \bigg( \frac{\partial \nu(\theta)}{\partial\theta} \bigg)^2,
\end{align}
by comparing \eqref{lower:bound:nonadditive:gauss} and \eqref{fisher:information:additive:gauss} it can be concluded that for all cases where
\begin{align}
\bar{\mu}_4(\theta)&\leq3,
\end{align}
a conservative system model $p(z;\theta)$ under an estimation theoretic perspective is the Gaussian model \eqref{model:Gaussian:generic:signal}.
\subsection{Information Loss of the Squaring Device}
Another interesting problem in statistical signal processing is to characterize the estimation theoretic quality of nonlinear receive and measurement systems. The Fisher information measure $F(\theta)$ is a rigorous tool which allows us to draw precise conclusions. However, depending on the nature of the nonlinearity, the exact calculation of the information measure $F(\theta)$ can become complicated. As an example for such a scenario, consider the problem of analyzing the intrinsic capability of a system with a squaring output (power sensor)
\begin{align}\label{definition:squaring:device}
Z=Y^2,
\end{align}
to infer the mean $\theta$ of a Gaussian input
\begin{align}
p(y;\theta)=\frac{1}{\sqrt{2\pi}} {\rm e}^{- \frac{(y-\theta)^2}{2} }
\label{unit:var:gauss}
\end{align}
with unit variance. In such a case, the system output $Z$ follows a non-central chi-squared distribution with a single degree of freedom parameterized by $\theta$. As the analytical description of the associated probability density function $p(z;\theta)$ includes a Bessel function, the characterization of the Fisher information $F(\theta)$ in compact analytical form is not trivial. We short-cut the derivation by using the approximation \eqref{fisher:information:bound:opt}. The first two moments of the output \eqref{definition:squaring:device} are found to be given by 
\begin{align}
\ex{Z}
&=\ex{\theta^2+2\theta W+W^2}\notag\\
&=\theta^2+1\notag\\
&=\mu_1(\theta),\\
\ex{(Z- \mu_1(\theta))^2}&=\ex{(\theta^2+2\theta W+W^2- \theta^2-1)^2}\notag\\
&=2(2\theta^2+1)\notag\\
&=\mu_2(\theta),
\end{align}
where we have introduced the auxiliary random variable
\begin{align}
W=Y-\theta.
\end{align}
The third output moment is
\begin{align}
\ex{(Z- \mu_1(\theta))^3}&=\ex{(\theta^2+2\theta W+W^2- \theta^2-1)^3}\notag\\
&=8(3\theta^2+1)\notag\\
&=\mu_3(\theta),
\end{align}
while the fourth moment is
\begin{align}
\ex{(Z- \mu_1(\theta))^4}&=\ex{(\theta^2+2\theta W+W^2- \theta^2-1)^4}\notag\\
&=12\big((2 \theta^2+1)^2 +4(4\theta^2+1)\big)\notag \\
&=\mu_4(\theta).
\end{align}
Therefore, the skewness and the kurtosis are
\begin{align}
\bar{\mu}_3(\theta)&={\mu_3(\theta)}{\mu_2^{-\frac{3}{2}}(\theta)}\notag\\
&=\frac{8(3\theta^2+1)}{2\sqrt{2}(2\theta^2+1)^{\frac{3}{2}}}\notag\\
&=\frac{2\sqrt{2}(3\theta^2+1)}{(2\theta^2+1)^{\frac{3}{2}}},\\
\bar{\mu}_4(\theta)&={\mu_4(\theta)}{\mu_2^{-2}(\theta)}\notag\\
&=\frac{12\big((2 \theta^2+1)^2 +4(4\theta^2+1)\big)}{4(2\theta^2+1)^2}\notag\\
&=\frac{12(4\theta^2+1)}{(2\theta^2+1)^2}+3.
\end{align}
With the derivatives 
\begin{align}
\frac{\partial \mu_1(\theta)}{\partial\theta}&=2\theta,\\ 
\frac{\partial \mu_2(\theta)}{\partial\theta}&=8\theta,
\end{align}
we obtain
\begin{align}
\beta^\star(\theta)&=\frac{\frac{\partial \mu_1(\theta)}{\partial\theta} \sqrt{\mu_2(\theta)}\bar{\mu}_3 (\theta) - \frac{\partial \mu_2(\theta)}{\partial\theta}}
{ \frac{\partial \mu_2(\theta)}{\partial\theta} \bar{\mu}_3 (\theta) -\frac{\partial \mu_1(\theta)}{\partial\theta} \sqrt{\mu_2(\theta)} (\bar{\mu}_4 (\theta)-1)}\notag\\
&= -\frac{  \theta^2\sqrt{2}\sqrt{(2 \theta^2+1)}}
{  ( 4 \theta^4 + 16\theta^2+3)}
\end{align}
and the approximation is finally given by
\begin{align}
S(\theta)&= \frac{1}{{\mu_2(\theta)}} \frac{\Big(\frac{\partial \mu_1(\theta)}{\partial\theta} + \frac{\beta^\star(\theta)}{\sqrt{\mu_2(\theta)}} \frac{\partial \mu_2(\theta)}{\partial\theta} \Big)^2}{1+ 2 \beta^\star(\theta) \bar{\mu}_3(\theta) + {\beta^\star}^2(\theta) (\bar{\mu}_4(\theta)-1)}\notag\\
&= \frac{2\theta^2 \big( 4 \theta^4 + 12\theta^2+3 \big)^2}{ \big( 4 \theta^4 + 12\theta^2+3 \big)  \big( 8\theta^6 +24\theta^4 + 18\theta^2+3 \big)  }\notag\\
&= \frac{2\theta^2 \big( 4 \theta^4 + 12\theta^2+3 \big)}{ \big( 8\theta^6 +24\theta^4 + 18\theta^2+3 \big)  }.
\end{align}
Fig. \ref{Squaring_Loss} depicts the conservative approximation
\begin{align}\label{approximative:loss:nonlinearity}
\tilde{\chi}(\theta)=\frac{S_Z(\theta)}{F_Y(\theta)}
\end{align}
of the information loss \eqref{exact:loss:nonlinearity} when squaring the random input variable $Y$. Note that Fig. \ref{Squaring_Loss} indicates that for small values of $\theta$ the squaring operation results in a strong degradation of the estimation capability.
\pgfplotsset{legend style={rounded corners=2pt,nodes=right}}
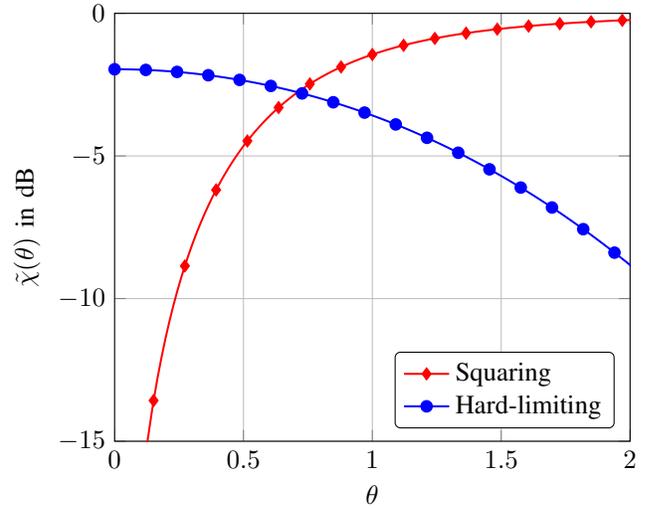
\begin{figure}[!htbp]
\begin{tikzpicture}[scale=1.0]

  	\begin{axis}[ylabel=$\tilde{\chi}(\theta)\text{ in dB}$,
  			xlabel=$\theta$,
			grid,
			ymin=-15,
			ymax=0,
			xmin=0,
			xmax=2,
			legend pos=south east]
			
    			\addplot[red, style=solid, line width=0.75pt,smooth, every mark/.append style={solid, fill=red}, mark=diamond*, mark repeat=4] table[x index=0, y index=1]{SquaringLoss.txt};
			\addlegendentry{Squaring}
			
			\addplot[blue, style=solid, line width=0.75pt,smooth, every mark/.append style={solid, fill=blue}, mark=otimes*, mark repeat=4] table[x index=0, y index=1]{HardLimiterLoss.txt};
			\addlegendentry{Hard-limiting}

\end{axis}	
\end{tikzpicture}
\caption{Performance Loss of Two Nonlinear Systems.}
\label{Squaring_Loss}
\end{figure}
As a comparison, the corresponding loss for a symmetric hard-limiter \cite{Vleck66}
\begin{align}\label{definition:hard:limiter}
Z=\operatorname{sign}{(Y)}
\end{align}
is visualized in Fig. \ref{Squaring_Loss}. Note that for hard-limiting the Gaussian model \eqref{unit:var:gauss}, it was shown in \cite{Stein14} that \eqref{eq:bound:unopt} forms a tight lower bound for the Fisher information measure. It can be observed that for small values of $\theta$, the information about the algebraic sign (hard-limiting) of the system input $Y$ conveys more information about the input mean $\theta$ than the amplitude (squaring). For $\theta \geq 0.75$, the situation changes as the statistics of the hard-limiter output vary slower with the parameter $\theta$ and therefore the squaring receiver outperforms the hard-limiter when it comes to estimating the mean $\theta$ of the system input $Y$ from samples at the system output $Z$. Note that for the squaring device \eqref{definition:squaring:device}, Fig. \ref{Squaring_Loss} depicts a conservative approximation \eqref{approximative:loss:nonlinearity} of the exact squaring loss \eqref{exact:loss:nonlinearity}.
\subsection{Measuring the Inference Capability with a Smooth Limiter}
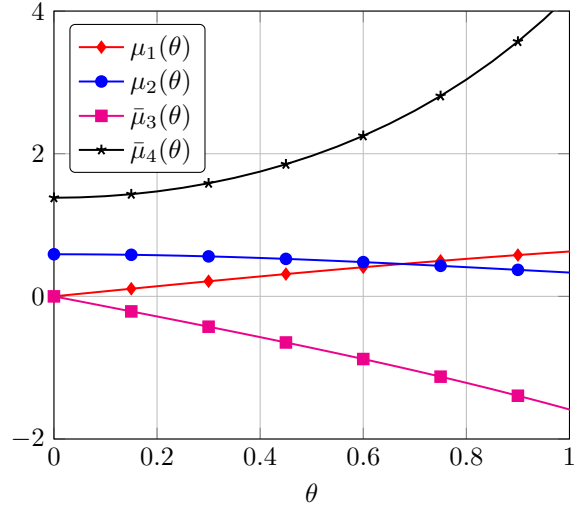
\begin{figure}[!htbp]
\begin{tikzpicture}[scale=1.0]

  	\begin{axis}[ylabel=$$,
  			xlabel=$\theta$,
			grid,
			ymin=-2.0,
			ymax=4.0,
			xmin=0,
			xmax=1,
			legend pos=north west]

			\addplot[red, style=solid, line width=0.75pt, every mark/.append style={solid, fill=red}, mark=diamond*, mark repeat=3] table[x index=0, y index=3]{SoftLimiterMean_0p5.txt};
			\addlegendentry{$\mu_1(\theta)$}
			
			\addplot[blue, style=solid, line width=0.75pt, every mark/.append style={solid, fill=blue}, mark=otimes*, mark repeat=3] table[x index=0, y index=4]{SoftLimiterMean_0p5.txt};
			\addlegendentry{$\mu_2(\theta)$}
			
			\addplot[magenta, style=solid, line width=0.75pt, every mark/.append style={solid, fill=magenta}, mark=square*, mark repeat=3] table[x index=0, y index=5]{SoftLimiterMean_0p5.txt};
			\addlegendentry{$\bar{\mu}_3(\theta)$}
			
			\addplot[black, style=solid, line width=0.75pt,every mark/.append style={solid, fill=black}, mark=star, mark repeat=3] table[x index=0, y index=6]{SoftLimiterMean_0p5.txt};
			\addlegendentry{$\bar{\mu}_4(\theta)$}
		
\end{axis}	
\end{tikzpicture}
\caption{Measured Moments ($\zeta=0.5$) of the Smooth Limiter.}
\label{fig:softlimiter:moments}
\end{figure}

\begin{figure}[!htbp]
\begin{tikzpicture}[scale=1.0]

  	\begin{axis}[ylabel=$$,
  			xlabel=$\theta$,
			grid,
			ymin=-1.0,
			ymax=1.0,
			xmin=0,
			xmax=1,
			legend pos=south west]

			\addplot[red, style=solid, line width=0.75pt, every mark/.append style={solid, fill=red}, mark=diamond*, mark repeat=3] table[x index=0, y index=7]{SoftLimiterMean_0p5.txt};
			\addlegendentry{$\frac{\partial\mu_1(\theta)}{\partial \theta}$}
			
			\addplot[blue, style=solid, line width=0.75pt, every mark/.append style={solid, fill=blue}, mark=otimes*, mark repeat=3] table[x index=0, y index=8]{SoftLimiterMean_0p5.txt};
			\addlegendentry{$\frac{\partial\mu_2(\theta)}{\partial \theta}$}
					
\end{axis}	
\end{tikzpicture}
\caption{Measured Derivatives ($\zeta=0.5$) of the Smooth Limiter.}
\label{fig:softlimiter:derivatives}
\end{figure}
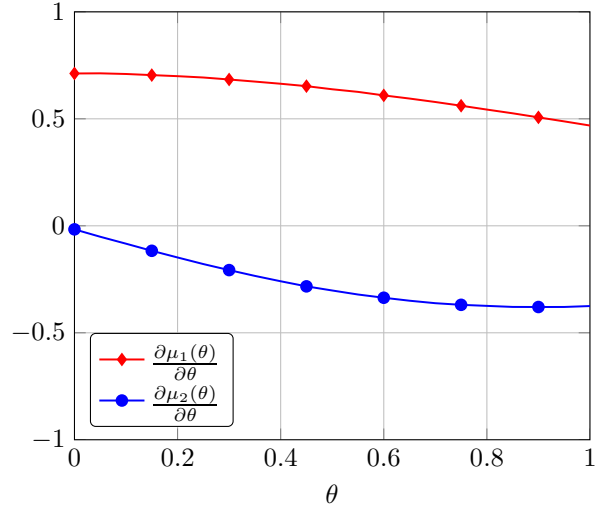

\begin{figure}[!htbp]
\begin{tikzpicture}[scale=1.0]

  	\begin{axis}[ylabel=$\tilde{\chi}(\theta)\text{ in dB}$,
  			xlabel=$\theta$,
			grid,
			ymin=-5,
			ymax=0,
			xmin=0,
			xmax=1,
			legend pos=south west]
			
			\addplot[black, style=solid, line width=0.75pt, every mark/.append style={solid, fill=black}, mark=diamond*, mark repeat=3] table[x index=0, y index=2]{SoftLimiterMean_1.txt};
			\addlegendentry{$\zeta=1.00$}
			
			\addplot[blue, style=solid, line width=0.75pt, every mark/.append style={solid, fill=blue}, mark=otimes*, mark repeat=3] table[x index=0, y index=2]{SoftLimiterMean_0p75.txt};
			\addlegendentry{$\zeta=0.75$}
			
			\addplot[green, style=solid, line width=0.75pt, every mark/.append style={solid, fill=green}, mark=square*, mark repeat=3] table[x index=0, y index=2]{SoftLimiterMean_0p5.txt};
			\addlegendentry{$\zeta=0.50$}
			
			\addplot[magenta, style=solid, line width=0.75pt,smooth,every mark/.append style={solid, fill=magenta}, mark=star, mark repeat=3] table[x index=0, y index=2]{SoftLimiterMean_0p25.txt};
			\addlegendentry{$\zeta=0.25$}
			
			\addplot[red, style=solid, line width=0.75pt,smooth,every mark/.append style={solid, fill=red}, mark=triangle*, mark repeat=3] table[x index=0, y index=2]{SoftLimiterMean_0p1.txt};
			\addlegendentry{$\zeta=0.10$}
			
			\addplot[black, style=dotted, line width=0.75pt,smooth] table[x index=0, y index=1]{HardLimiterLoss.txt};
			\addlegendentry{$ \zeta{\to} 0.00$}
					
\end{axis}	 
\end{tikzpicture}
\caption{Measured Information Loss of the Smooth Limiter.}
\label{fig:softlimiter:loss}
\end{figure}
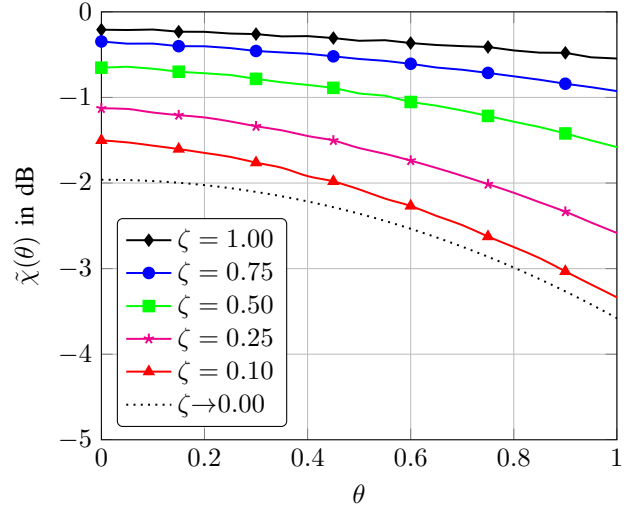

A situation that is often encountered in practice is that the analytical characterization of  the system model $p(z;\theta)$ is difficult. If the appropriate system model $p(z;\theta)$ is unknown \cite{Berisha15}, the direct consultation of an analytical tool like the Fisher information measure $F(\theta)$ becomes impossible. However, in such a situation, an information bound like $S(\theta)$ allows us to numerically approximate the information measure $F(\theta)$ at low-complexity. To this end, the moments of the system output $Z$ are measured in a calibrated setup, where the parameter $\theta$ can be controlled, or determined by Monte-Carlo simulations. We demonstrate this validation technique by using a smooth limiter model, i.e., the system input $Y$ is transformed by
\begin{align}
Z&=\sqrt{\frac{2}{\pi\zeta^2}}\int_{0}^{Y} \exp{-\frac{u^2}{2\zeta^2}} {\rm d}u\notag\\
&=\erf{\frac{Y}{\sqrt{2\zeta^2}}},\label{system:soft:limiter}
\end{align}
where $\zeta\in\fieldR$ is a constant model parameter and 
\begin{align}
\erf{x}=\frac{2}{\sqrt{\pi}}\int_{0}^{x} \exp{-t^2} {\rm d}t
\end{align}
is the error function. This nonlinear model \cite{Baum57} can be used in order to characterize saturation effects in analog system components like amplifiers \cite{Pawula71} \cite{Cann80}. 
\begin{figure}[!htbp]
\begin{tikzpicture}[scale=1.0]

  	\begin{axis}[ylabel=$z$,
  			xlabel=$y$,
			grid,
			ymin=-1.0,
			ymax=1.0,
			xmin=-1,
			xmax=1,
			legend pos=south east]

			\addplot[red, style=solid, line width=0.75pt,smooth, every mark/.append style={solid, fill=red}, mark=diamond*, mark repeat=3] table[x index=0, y index=5]{SoftLimiter_IO.txt};
			\addlegendentry{$\zeta=0.10$}
			
			\addplot[magenta, style=solid, line width=0.75pt,smooth, every mark/.append style={solid, fill=magenta}, mark=otimes*, mark repeat=3] table[x index=0, y index=4]{SoftLimiter_IO.txt};
			\addlegendentry{$\zeta=0.25$}
			
			\addplot[green, style=solid, line width=0.75pt,smooth,every mark/.append style={solid, fill=green}, mark=square*, mark repeat=3] table[x index=0, y index=3]{SoftLimiter_IO.txt};
			\addlegendentry{$\zeta=0.50$}
			
			\addplot[blue, style=solid, line width=0.75pt,smooth,every mark/.append style={solid, fill=blue}, mark=star, mark repeat=3] table[x index=0, y index=2]{SoftLimiter_IO.txt};
			\addlegendentry{$\zeta=0.75$}
			
			\addplot[black, style=solid, line width=0.75pt,smooth,every mark/.append style={solid, fill=black}, mark=triangle*, mark repeat=3] table[x index=0, y index=1]{SoftLimiter_IO.txt};
			\addlegendentry{$\zeta=1.00$}
		
\end{axis}	
\end{tikzpicture}
\caption{Input-to-Output Relation of the Smooth Limiter.}
\label{fig:softlimiter:io}
\end{figure}
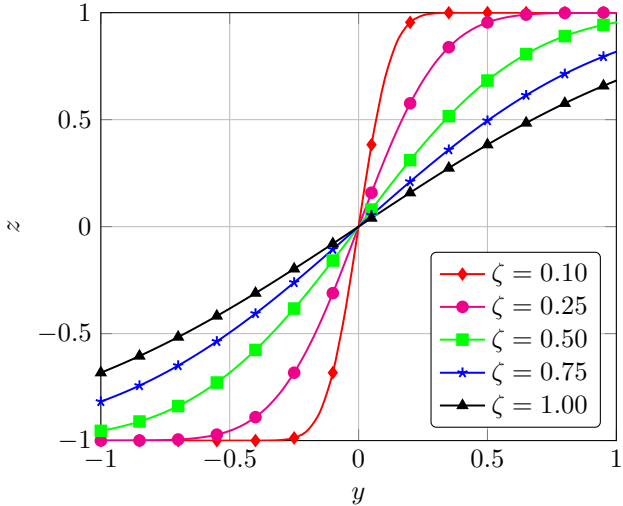
In Fig. \ref{fig:softlimiter:io}, the input-to-output mapping of the model (\ref{system:soft:limiter}) is depicted for different setups $\zeta$.
As input, we consider a Gaussian distribution with unit variance like in (\ref{unit:var:gauss}). The output mean $\mu_1(\theta)$, variance $\mu_2(\theta)$, skewness $\bar{\mu}_3(\theta)$, and kurtosis $\bar{\mu}_4(\theta)$ are measured by $10^9$ independent Monte-Carlo simulations of the nonlinear system output $Z$ for each considered value of the input mean $\theta$. The result is shown in Fig. \ref{fig:softlimiter:moments}. 
After numerically approximating the required derivatives $\frac{\partial \mu_1(\theta)}{\partial \theta}, \frac{\partial \mu_2(\theta)}{\partial \theta}$, which are depicted in Fig. \ref{fig:softlimiter:derivatives}, the approximation $S(\theta)$ is calculated. 
In Fig. \ref{fig:softlimiter:loss}, the measured information loss $\tilde{\chi}(\theta)$ of the smooth limiter model is shown, where the dotted line indicates the exact information loss ${\chi}(\theta)$ with a hard-limiter \eqref{definition:hard:limiter} (as depicted in Fig. \ref{Squaring_Loss}), which is equivalent to a smooth limiter with $\zeta\to0$.
\section{Conclusion}
We have established a generic and compact lower bound for the Fisher information measure. By various examples we have shown that the derived expression has the potential to provide a good approximation in a broad number of cases. This makes the presented information bound a versatile mathematical tool for a variety of problems encountered in the design and optimization of signal processing systems. Further, the pessimistic nature of the attained alternative information measure allows us to strengthen insights on worst-case noise and to generalize classical results on Gaussian system models which exhibit minimum Fisher information. Finally, we have outlined how to use the presented information bound in order to benchmark the estimation capability of physical measurement systems with output statistics of unknown analytical form.
\bibliographystyle{IEEEbib}

\end{document}